\documentclass[12pt]{article}
\usepackage{epsfig}\parskip 5pt plus 1pt
\newcommand{\be}{\begin{equation}}
\newcommand{\ee}{\end{equation}}
\newcommand{\bea}{\begin{eqnarray}}
\newcommand{\eea}{\end{eqnarray}}

\def\simge{\mathrel{%
   \rlap{\raise 0.511ex \hbox{$>$}}{\lower 0.511ex \hbox{$\sim$}}}}
\def\simle{\mathrel{
   \rlap{\raise 0.511ex \hbox{$<$}}{\lower 0.511ex \hbox{$\sim$}}}}

\begin{document}
\thispagestyle{empty}
\vspace*{1cm}
\begin{center}
{\Large{\bf Next generation long baseline experiments
  on the path to leptonic CP violation  } }\\
\vspace{.5cm}
P. Migliozzi$^{\rm a,}$\footnote{pasquale.migliozzi@cern.ch} and
F. Terranova $^{\rm b,}$\footnote{francesco.terranova@cern.ch} \\
\vspace*{1cm}
$^{\rm a}$ I.N.F.N., Sezione di Napoli, Naples, Italy \\
$^{\rm b}$ I.N.F.N., Laboratori Nazionali di Frascati,
Frascati (Rome), Italy 
\end{center}

\vspace{.3cm}
\begin{abstract}
\noindent
In this paper we quantify the trade-off between setups optimized to be
ancillary to Phase~II Superbeams or Neutrino Factories and experiments
tuned for maximal sensitivity to the subdominant terms of the neutrino
transition probability at the atmospheric scale (``maximum discovery
potential''). In particular, the $\theta_{13}$ sensitivity is computed
for both Phase~I superbeams (JHF-SK and NuMI~Off-Axis) and next generation
long baseline experiments (ICARUS, OPERA and MINOS).  It is shown that
Phase~I experiments cannot reach a sensitivity able to ground (or
discourage in a definitive manner) the building of Phase~II projects
and that, in case of null result and without a dedicated $\bar{\nu}$ run,
this capability is almost saturated by high energy beams like
CNGS, especially for high values of the ratio 
$\Delta m^2_{21}/|\Delta m^2_{31}|$.
\end{abstract}

\vspace*{\stretch{2}}
\begin{flushleft}
\end{flushleft}

\newpage

\section{Introduction}
\label{introduction}
The possibility to perform a CKM-like precision physics in the
leptonic sector employing terrestrial neutrino oscillation experiments
has been deeply debated in the last few years.  At present, the
occurrence of neutrino oscillations seems rather well established
\cite{evidence_osc,K2K,KAMLAND}. Current experimental evidence
indicates two hierarchical mass scale differences ($\Delta m^2_{21}
\ll |\Delta m^2_{32}| \simeq |\Delta m^2_{31}|$) driving,
respectively, the oscillations at the ``solar'' and ``atmospheric''
scale.  Moreover, the $\alpha \equiv \Delta m^2_{21}/ | \Delta
m^2_{31} |$ ratio is constrained by the LMA solution of the solar
neutrino puzzle to lie between ${\cal O}(0.1)$ and ${\cal
O}(0.01)$~\cite{fogli_lisi_kamland}.  If this scenario will be
confirmed after the completion of ongoing experiments (K2K~\cite{K2K},
KAMLAND~\cite{KAMLAND} and MiniBoone~\cite{MINIBOONE}) and next
generation long baseline projects (MINOS~\cite{MINOS},
ICARUS~\cite{ICARUS}, OPERA~\cite{OPERA}), terrestrial neutrino
experiments based on ``Superbeams'' (SB) or ``Neutrino Factories''
(NF) could be the ideal tool for precision measurements of the
PMNS~\cite{PMNS} leptonic mixing matrix and the discovery of leptonic CP
violation \cite{literature_sb_nf}. These experiments explore
subdominant effects in the neutrino transition probabilities at the
atmospheric scale which, in general, are suppressed by at least one
power of $\alpha$. Hence, the recent KAMLAND result places SB and NF
proposals on a firmer ground since guarantees that subdominant effects
will not be suppressed to an unobservable level ($\alpha \ll
10^{-2}$). This condition, however, is not enough to establish the physics
reach of SB/NF. As for the case of CKM physics, CP violating effects
depend on the size of the Jarlskog invariant \cite{jarlskog}. In the
standard parameterization \cite{pdg} of the PMNS matrix this
coefficient can be expressed as:

\begin{equation}
J \equiv s_{12} s_{23} s_{13} c_{12} c_{23} c_{13}^2 \sin \delta =
\frac{1}{8} \sin 2\theta_{12} \sin 2\theta_{23} \sin 2\theta_{13} c_{13} 
\sin \delta    
\end{equation}

\noindent where $s_{ij} \equiv \sin \theta_{ij}$ and $c_{ij} \equiv
\cos \theta_{ij}$. Differently from the quark case, the leptonic
Jarlskog invariant is enhanced by the large mixing angles
$\theta_{23}$ and $\theta_{12}$. On the other hand, due to the null
result of the CHOOZ~\cite{CHOOZ} and PALO VERDE~\cite{paloverde}
experiments, the full three-flavor mixing of neutrinos is still
unestablished and only upper limits on the $\sin^2 2 \theta_{13}$
parameter have been drawn ($\sin^2 2 \theta_{13}< {\cal
O}(10^{-1})$). Moreover, no theoretical inputs are available to
constrain the size of $\theta_{13}$ in a convincing manner, so that
its experimental determination is mandatory. The discovery of
$\theta_{13} \neq 0$ has not only a scientific relevance but also a
high practical value. The commissioning and running of an apparatus to
observe CP violation in the leptonic sector at the atmospheric scale
(e.g. JHF-Phase~II or a Neutrino Factory) is a major technical and
economical challenge; since most of its physics reach - in particular
the measurement of leptonic CP violation and the determination of
$U_{e3}$ in the PMNS matrix - depends crucially on the size
of $\theta_{13}$, the latter should be determined by ``Phase~I''
experiments (e.g. JHF-SK~\cite{JHF} or NuMI~Off-Axis~\cite{numioa})
tuned to maximize their $\theta_{13}$ sensitivity.  Otherwise, the
physics case of SB/NF should be drawn independently of their PMNS
reach. This is marginally possible for JHF-Phase~II (proton decay with
HyperK) but rather unrealistic for NF.  The physics case of
Phase~I experiments is very appealing due to their unprecedented
precision in the determination of the parameters leading the
oscillations at the atmospheric scale ($\theta_{23}$ and $|\Delta
m^2_{31}|$) and their significant discovery potential for high values
of $\theta_{13}$.  On the other hand, the sensitivity of Phase~I
experiments to $\theta_{13}$ has been questioned since a significant
deterioration is expected once we account for our complete ignorance
of the leptonic CP phase, the sign of $\Delta m^2_{31}$
and the $\theta_{23}$ ambiguity~\cite{lindner,kajita}.
In this context, the advantage of a ``pure'' $\theta_{13}$
measurement has been put forward especially in connection with new
reactor experiments \cite{minakata}.  

In this letter we quantify the
trade-off between a setup optimized to be ancillary with respect the
SB/NF project (maximum $\theta_{13}$ sensitivity) and one highly
sensitive to the subdominant terms of the transition probability
(maximum discovery potential). In particular, we challenge the claim
that a Phase~I experiment can reach a sensitivity able to ground (or
discourage in a definitive manner) the building of  SB/NF
and show that, in case of null results and without a dedicated
$\bar{\nu}$ run,  this capability is almost saturated
by first generation long baseline experiments like CNGS.

\section{Oscillation probabilities}
\label{osc_prob}

The next generation long baseline experiments (MINOS, ICARUS and
OPERA) and Phase~I experiments (JHF-SK and NuMI~Off-Axis) employ
baselines in the 300-700~km range. In most of the cases, the neutrino
energy is optimized to maximize the oscillation probability at the
atmospheric scale for the corresponding baseline ($\langle E_\nu
\rangle \simeq 0.7-3$~GeV). The CNGS experiments, however, make use of
a high energy beam, well beyond the kinematic threshold for $\tau$
production ($\langle E_\nu \rangle \simeq 17$~GeV).  The main
parameters for the setups under consideration are listed in
Table~\ref{tab:list_parameter}. In all cases the subleading
oscillations at the solar scale are suppressed by at least one order
of magnitude compared with the atmospheric ones. Hence, oscillation
probabilities can be expanded in the small parameters $\alpha$ and
$\sin 2\theta_{13}$.  The inclusion of matter effects is simplified
here, since the earth density can be considered constant along
baselines shorter than $\sim$1000~km. In particular, the $\nu_\mu
\rightarrow \nu_e$ oscillation probability can be expressed
as~\cite{cervera,freund}:

\begin{eqnarray}
P_{\nu_\mu \rightarrow \nu_e} & \simeq & \sin^2 2\theta_{13} \, \sin^2
\theta_{23} \frac{\sin^2[(1- \hat{A}){\Delta}]}{(1-\hat{A})^2}
\nonumber \\ & - & \alpha \sin 2\theta_{13} \, \xi \sin \delta
\sin({\Delta}) \frac{\sin(\hat{A}{\Delta})}{\hat{A}}
\frac{\sin[(1-\hat{A}){\Delta}]}{(1-\hat{A})} \nonumber \\ &+& \alpha
\sin 2\theta_{13} \, \xi \cos \delta \cos({\Delta})
\frac{\sin(\hat{A}{\Delta})}{\hat{A}} \frac{\sin[(1-\hat{A}){\Delta}]}
{(1-\hat{A})} \nonumber \\ &+& \alpha^2 \, \cos^2 \theta_{23} \sin^2
2\theta_{12} \frac{\sin^2(\hat{A}{\Delta})}{\hat{A}^2} \nonumber \\ &
\equiv & O_1 \ + \ O_2(\delta) \ + \ O_3(\delta) \ + \ O_4 \ \ .
\label{equ:probmatter}
\end{eqnarray}

\noindent In this formula $\Delta \equiv \Delta m_{31}^2 L/(4 E)$ and
the terms contributing to the Jarlskog invariant are split into the
small parameter $\sin 2\theta_{13}$, the ${\cal O}(1)$ term $\xi
\equiv \cos\theta_{13} \, \sin 2\theta_{12} \, \sin 2\theta_{23}$ and
the CP term $\sin \delta$; $\hat{A} \equiv 2 \sqrt{2} G_F n_e E/\Delta
m_{31}^2$ with $G_F$ the Fermi coupling constant and $n_e$ the
electron density in matter. Note that the sign of $\hat{A}$ depends on
the sign of $\Delta m_{31}^2$ which is positive (negative) for normal
(inverted) hierarchy of neutrino masses. The dominant contributions
among the four terms $O_1 \ldots O_4$ of Eq.~\ref{equ:probmatter} are
determined by the choice of $L$ and $E$. In the following, if not
stated explicitly, we assume the present best fits for the solar and
atmospheric parameters ($\Delta m^2_{21}=7.3 \times 10^{-5}$~eV$^2$,
$\sin^2 2\theta_{12}=0.8$, $\Delta m^2_{31}=2.5 \times
10^{-3}$~eV$^2$, $\sin^2
2\theta_{23}=1$)~\cite{fogli_lisi_kamland}~\footnote{ For
$\theta_{23}\ne \pi/4$ other degenerate solutions appear at
$\theta_{23}' = \pi/2 -\theta_{23}$ \cite{barger_theta_12_amb}.}.

\begin{table}[hbtp]
\begin{center}
\begin{tabular}{l|c|c|c|c|c}
\hline
\hline
                               & JHF-SK & NuMI-OA & MINOS & ICARUS & OPERA \\ \hline
Baseline~(km)                   & 295    & 712     & 735   & 732    & 732 \\
Mean energy~(GeV)               & 0.76   & 2.22    & 3     & 17     & 17  \\
Exposure~(kton$\times$years) & 22.5$\times$5 & 17$\times$5 
                             & 5.4$\times$2 & 2.4$\times$5 & 1.7$\times$5   \\
L/E (km/GeV)               & 388       & 321     & 245   & 43     & 43 \\
\hline
\end{tabular}
\caption{Main parameters of the Phase~I and
long baseline experiments.}
\label{tab:list_parameter}
\end{center}
\end{table}

\subsection*{JHF-SK}

\begin{figure}[htbp]
\centering
\epsfig{file=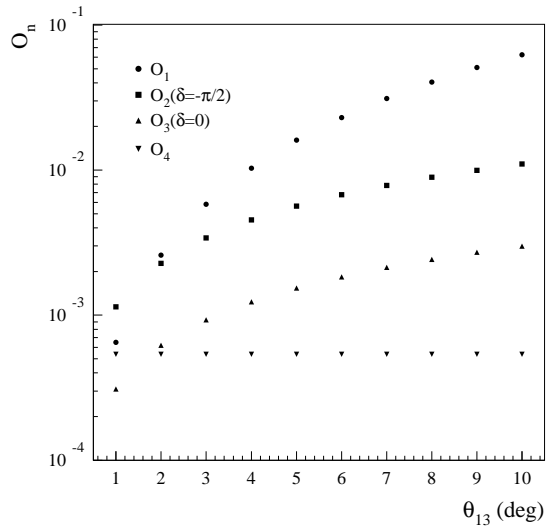,width=8cm}
\caption{Contribution of the $O_1 \ldots O_4$ terms to the oscillation
probability in the JHF-SK scenario.}
\label{fig:terms}
\end{figure}

JHF-SK has been tuned to maximize the discovery potential and
subdominant contributions depending on the CP phase are
enhanced. Given its short baseline 
matter effects represent a small correction to the
oscillation probability ($\hat{A} \simeq 5
\times 10^{-2}$).  Assuming an average neutrino energy of 0.76~GeV,
the following hierarchy among the terms of Eq.~\ref{equ:probmatter} is
obtained:

\begin{eqnarray}
P_{\nu_\mu \rightarrow \nu_e} & \simeq &  \sin^2 2\theta_{13} A_1
- \sin \delta  \sin 2\theta_{13} \ \alpha A_2
+ \cos \delta  \sin 2\theta_{13} \ \alpha \cos (\Delta) \ A_3 +
\alpha^2 A_4
\label{equ:prob_jhf}
\end{eqnarray}

\noindent where the $A_i$ ($i=1, \ldots 4$) 
coefficients are ${\cal O}(1)$. The actual values
of the terms contributing to  Eq.~\ref{equ:probmatter} are shown in 
Fig.~\ref{fig:terms}. Here, the $\delta$-depending terms 
$O_2$ and $O_3$ are computed at maximum amplitude, $O_2=O_2(\delta=-\pi/2)$
and $O_3=O_3(\delta=0)$, to illustrate the impact of assuming
complete ignorance on $\delta$ in the extraction of $\sin^2 2\theta_{13}$.
Of course, in the oscillation probability formula 
when $O_2$ ($O_3$) is maximal, i.e. $\delta=-\pi/2$ ($\delta=0$),
the other coefficient $O_3$ ($O_2$) is zero.
For $\sin^2 2 \theta_{13}$ sufficiently high:

\begin{equation}
P_{\nu_\mu \rightarrow \nu_e} \simeq  
\sin 2\theta_{13}  ( \sin 2\theta_{13} A_1 - \sin \delta \ \alpha \ A_2)
\label{equ:prob_jhf_approx}
\end{equation}

\noindent
Eq.~\ref{equ:prob_jhf} and ~\ref{equ:prob_jhf_approx} 
show that the deterioration of the $\sin^2 2\theta_{13}$ sensitivity
coming from the $\theta_{13}$-$\delta$ ambiguity~\cite{burguet-castell}
is strictly connected
to the mass scale ratio $\alpha$. Hence, the maximum sensitivity
is achieved in the limit $\alpha \rightarrow 0$ which corresponds to
minimum sensitivity to the subdominant terms of 
$P_{\nu_\mu \rightarrow \nu_e}$ (minimum discovery potential). Clearly,
this contradictory request is at the origin of the conflict between 
setups ancillary to SB/NF and experiments able to explore a significant
fraction of the PMNS parameter space. For JHF-SK, the deterioration effect
becomes sizable already at $\theta_{13} \sim 3^\circ$ 
(see Fig.~\ref{fig:terms}). 
Values of $\alpha$ higher than
the ones assumed in Fig.~\ref{fig:terms} ($\alpha \simeq 0.03$) imply earlier
appearance of the ($\theta_{13}-\delta$) deterioration effect.  

Note that Eq.~\ref{equ:prob_jhf} cannot be used in a straightforward manner
to extract the actual  sensitivity  of a Phase~I experiment. The 
signal rate is: 

\begin{equation}
S \equiv Y \ A \int dE \ \Phi(E) \  
P_{\nu_\mu \rightarrow \nu_e}(E) \ \sigma(E) \ \epsilon(E) 
\label{equ:signalrate}
\end{equation}

\noindent where $\Phi(E)$ is the $\nu_\mu$ flux at the surface of the
detector, $A$ is proportional to the detector 
mass, $Y$ are the years of data taking and $\sigma(E) \cdot \epsilon(E)$ is 
the production cross-section weighted with the detection efficiency
for the $\nu_e$~CC final state. The signal rate
is proportional to  $P_{\nu_\mu \rightarrow \nu_e}(\langle E \rangle)$
only in the narrow band limit $\Phi(E) \rightarrow 
\delta(E- \langle E \rangle)$. The minimum accessible probability $P_{min}$
depends on the background rate and  it has to be computed through a
full simulation. We further address this issue in Sec.~\ref{numerical_calc}.
Finally, note that for $P_{min}$ sufficiently low (``Phase~II'' measurements),
setups can be envisaged to lift explicitly the $\theta_{13}-\delta$ 
ambiguity, e.g. combining different baselines \cite{diff_baselines} or
different oscillation channels \cite{donini} or building a single baseline
experiment with a detector capable of observing more than one 
oscillation peak \cite{single_baseline}.

\subsection*{NuMI~Off-Axis and MINOS}

The NuMI~Off-Axis proposal envisages the possibility of getting a very narrow
$\nu_\mu$ beam placing a dedicated detector for $\nu_e$ appearance
(20~kton, low-Z calorimeter) at an angle of $\sim 0.7^\circ$ with respect to the
present NuMI axis. Again, L and $E$ are tuned
close to the first oscillation maximum at the atmospheric scale. A 
significant reduction of the background coming from 
$\nu_\mu$~NC with $\pi^0$ production
can be reached, compared with the MINOS setup, thanks to the suppression
of the high energy tail of the $\nu_\mu$ beam. Once more, the terms
contributing to Eq.~\ref{equ:probmatter} keeps the form of 
Eq.~\ref{equ:prob_jhf} with $A_i$ ($i=1, \ldots 4$)  ranging between 0.4 and 
0.6. However, both MINOS and NuMI~Off-Axis employ a baseline of
$\simeq 700$~km and matter effects are sizable 
($\hat{A} \simeq 0.2$) in this regime since they modify the size of the 
leading term $A_1$. As a consequence, these setups offer a significant
sensitivity to the sign of $\Delta m_{31}^2$. On the other hand, if 
it is not possible to disentangle the sign($\Delta m_{31}^2$) 
degeneracy from the effect proportional to $\sin^2 2\theta_{13}$, an additional
source of deterioration of the $\sin^2 2\theta_{13}$ sensitivity
appears. In principle, it could be possible to re-tune NuMI~Off-Axis
releasing the condition $\Delta \simeq \pi/2$ and, hence, modifying
the relative weights of the $A_{i}$ coefficients. In this scenario, 
NuMI would be complementary to  JHF-SK since the former would 
lower its bare $\sin^2 2\theta_{13}$ sensitivity allowing the
latter to relieve the ($\delta-\theta_{13}$) 
deterioration discussed above. This possibility 
and the overall improvement in the PMNS reach of the synergic JHF/NuMI physics
programme has been discussed in details in \cite{lindner,minakata} 
and will not be further considered here. 

\subsection*{CNGS}

\begin{figure}[htbp]
\centering
\epsfig{file=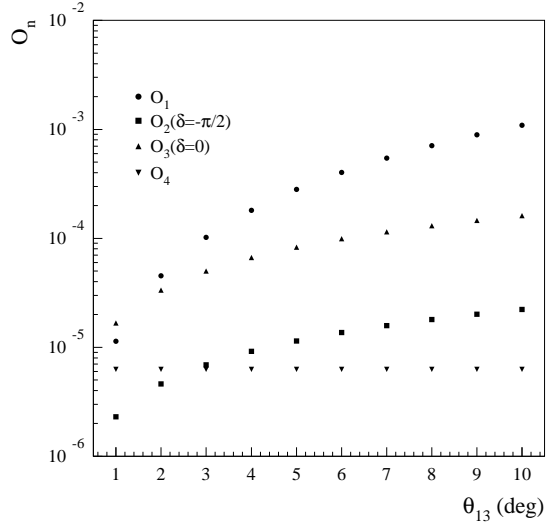,width=8cm}
\caption{Contribution of the $O_1 \ldots O_4$ terms to the oscillation
probability at CNGS.}
\label{fig:terms_cngs}
\end{figure}

The CNGS beam has been tuned to reach maximum sensitivity to the $\nu_\tau$
appearance channel. To overcome the limitation of the high threshold 
for $\tau$  production, the condition $\Delta \simeq \pi/2$ has been
given up and $\Delta_{CNGS} \simeq {\cal O}(10^{-1})$. As a consequence, 
the oscillation  probability is suppressed by the dumping term 
$\Delta^2 \simeq {\cal O}(10^{-2})$

\begin{equation}
P(\nu_\mu \rightarrow \nu_\tau) \simeq \cos^4 \theta_{13} 
\sin^2 2 \theta_{23} \Delta^2
\end{equation}

\noindent but the event rate profits of the high $\nu_\tau$-CC
cross-section.  The same dumping factor limits the search for $\nu_\mu
\rightarrow \nu_e$.  Again, this loss of signal events is partially
compensated by the linear rise of the $\nu_e$-CC cross-section and by
the high granularity of the corresponding detectors tuned for
$\nu_\tau$ (in particular $\tau \rightarrow e$) appearance and hence,
extremely effective in suppressing the NC($\pi^0$) and $\nu_\mu
\rightarrow \nu_\tau \rightarrow \tau (\rightarrow e) X$
background. It has been shown \cite{komatsu} that ICARUS and OPERA
combined could explore the region down to $\sin^2 2\theta_{13} \sim
0.025$ at $|\Delta m_{31}^2|=2.5\times10^{-3}$~eV$^2$ and assuming
$6.75\times10^{19}$~pot/year ($\sin^2 2\theta_{13} < 0.03$ for ICARUS
and $\sin^2 2\theta_{13} < 0.05$ for OPERA separately).  The analysis
is dominated by the statistical fluctuations of the $\nu_e$ beam
contamination from $K_{e3}$ decays and, for higher exposure time, by
the systematics uncertainty on its overall normalization~\footnote{The
CNGS physics programme does not foresee the construction of a near
detector.} (see Fig.~\ref{fig:years} in Section~\ref{numerical_calc}).
However, this analysis does not include the deterioration effect
coming from the CP phase and the sign of $\Delta m_{31}^2$.  In
principle, matter effects should be even higher than NuMI because
$\hat{A}$ grows linearly with $E$ and the two setups have the same
baseline ($\hat{A} \simeq 1.6$).  However, since $|(1-\hat{A})\Delta|
\ll 1$, we get:

\begin{equation}
\frac{\sin^2[(1- \hat{A}){\Delta}]}{(1-\hat{A})^2} \simeq \Delta^2 \ 
\end{equation}

\noindent and the leading  term $A_1$ turns out to be unaffected by the sign
of $\Delta m^2_{31}$. Eq.~\ref{equ:probmatter} reads now:

\begin{equation}
P_{\nu_\mu \rightarrow \nu_e} \simeq \left[ \sin^2 2\theta_{13} \  \ A_1 -
\sin \delta \ \sin 2\theta_{13} \ \alpha \ \Delta \ A_2 +
\cos \delta \ \sin 2\theta_{13} \ \alpha \ A_3 +
\alpha^2 \ A_4 \right] \Delta^2
\end{equation}
 
\noindent and, again, $A_1 \ldots A_4$ are ${\cal O}(1)$ coefficients,
albeit different from the ones of Eq.~\ref{equ:prob_jhf}. 
The values of the terms contributing to  Eq.~\ref{equ:probmatter} 
at CNGS are shown in  Fig.~\ref{fig:terms_cngs}.
The overall scale is suppressed by $\Delta^2$ and
the role of the $O_2$ and $O_3$ terms  are exchanged w.r.t. JHF
due to the different size of $\sin \Delta$ and $\cos \Delta$.
Here, $O_3$ is responsible for the ($\delta-\theta_{13}$) deterioration effect
which is sizable in the same $\theta_{13}$ region as for JHF. 
Note, however, that $O_3$ ($O_2$) is odd (even) 
under the transformation $\Delta m^2_{31}
\rightarrow - \Delta m^2_{31}$; so, at CNGS, going from normal to inverted
hierarchy has the same effect of performing
a $\delta \rightarrow \pi - \delta$ transformation in the CP phase.
Note also that both JHF-SK and CNGS see a deterioration of their sensitivity 
to  $\theta_{13}$ starting from 
the $3^\circ$ region (or before for higher $\alpha$).   

\section{Numerical calculations}
\label{numerical_calc}

Analyses of the $\nu_\mu \rightarrow \nu_e$ channel in the leading
order approximation $P(\nu_\mu \rightarrow \nu_e)\simeq O_1$ 
have been published by the collaborations involved in the Phase~I and next
generation long baseline experiments. Hence,
it is possible to make a reliable estimation of the actual sensitivities
side-stepping the full simulation of the various setups. 
The condition that excludes a point $(\sin^2 2\theta_{13},\Delta m^2_{31})$
of the parameter space at a given confidence level, once
the null hypothesis $\sin^2 2\theta_{13}=0$ has been experimentally observed
and a given value of $\delta$ is assumed, is

\begin{equation}
 \chi^2(\sin^2 2\theta_{13},\Delta m^2_{31} )  > \zeta 
\label{equ:exclusion}
\end{equation}

\noindent where

\begin{eqnarray}
\chi^2(\sin^2 2\theta_{13},\Delta m^2_{31} )
\ \equiv \ \left( \frac{R^{th}-R^{obs}}{\sigma} \right)^2  \ = \nonumber \\ 
\frac{ \left[ (S(\sin^2 2\theta_{13},\Delta m^2_{31})+B) 
\ - \ (S_{null}+B ) 
\right]^2} { S_{null}+B+
\eta^2 B^2 } > \zeta 
\label{equ:chi2}
\end{eqnarray}

\noindent In this formula, which holds in Gaussian approximation,
$R^{th}$ is the expected $\nu_e$ rate for the current value 
of ($\sin^2 2\theta_{13},\Delta m^2_{31}$) and $R^{obs}$ is the
rate corresponding to the null hypothesis. Eq.~\ref{equ:probmatter} 
shows that the null hypothesis is independent of the CP phase
and the sign of $\Delta m^2_{31}$
and depends only on $|\Delta m^2_{31}|$. $S$ and $B$ represent the signal and 
background rate, $\eta$ is the systematic uncertainty on the
background normalization and $\zeta$ is a constant depending on the confidence
level ($\zeta = 4.6$ for 90\%~CL contours). 
If the $\tau \rightarrow e$ contamination is negligible
w.r.t. the NC($\pi^0$) and the $\nu_e$ contamination from the beam
in the $\Delta m^2_{31}$ region of interest, 
$B$ is independent of the oscillation parameters. 
Dropping the $\Delta m^2_{31}$ dependence, the
minimum value of $\sin^2 2\theta_{13}$ excluded by the experiment 
is the one fulfilling:

\begin{equation}
 \frac{ \left[ (S(\sin^2 2\theta_{13})+B) 
\ - \ (S_{null}+B ) 
\right]^2} { S_{null}+B+
\eta^2 B^2 } = \zeta
\label{equ:min_excluded}
\end{equation}

The expected signal rate in Eq.~\ref{equ:chi2} can be written
(see Eq.~\ref{equ:signalrate}) as:

\begin{equation}
S (\sin^2 2\theta_{13},\delta) \equiv Y \ A \int dE \ \Phi(E) \  
P_{\nu_\mu \rightarrow \nu_e}(\sin^2 2\theta_{13},\delta,E) 
\ \sigma(E) \ \epsilon(E) 
\label{equ:signalrate2}
\end{equation}

\noindent and in the narrow beam approximation 
($\bar{E} \equiv \langle E \rangle$ )

\begin{equation}
S (\sin^2 2\theta_{13},\delta) \ = \ Y \gamma  
P_{\nu_\mu \rightarrow \nu_e}(\sin^2 2\theta_{13},\delta,\bar{E} )
\end{equation}

\noindent
where 

\begin{equation}
\gamma \equiv  A \ \Phi(\bar{E})   
\ \sigma(\bar{E}) \ \epsilon(\bar{E}) 
\end{equation}

\noindent Similarly, $S_{null}=Y \gamma 
P_{\nu_\mu \rightarrow \nu_e}
(\sin^2 2\theta_{13}=0,\bar{E} ) \equiv Y\gamma P_{null}(\bar{E}) $ 
and $B\equiv Y \beta$, being $\beta$ the background rate per year.
Now Eq.~\ref{equ:chi2} reads:

\begin{eqnarray}
P(\sin^2 2\theta_{13},\delta,\bar{E} ) & > & 
P_{null}(\bar{E})   + \sqrt{\zeta} 
\left\{ \frac{\beta}{Y\gamma^2} + \frac{\eta^2 \beta^2}{\gamma^2}+ 
\frac{P_{null}(\bar{E}) }{Y\gamma} \right\} ^{1/2}
\simeq \nonumber \\
& & P_{null}(\bar{E})  + \sqrt{\zeta} 
\left\{ \frac{\beta}{Y\gamma^2} + \frac{\eta^2 
\beta^2}{\gamma^2} \right\} ^{1/2} 
\label{equ:final_inequality}
\end{eqnarray}

\begin{figure}[htbp]
\centering
\epsfig{file=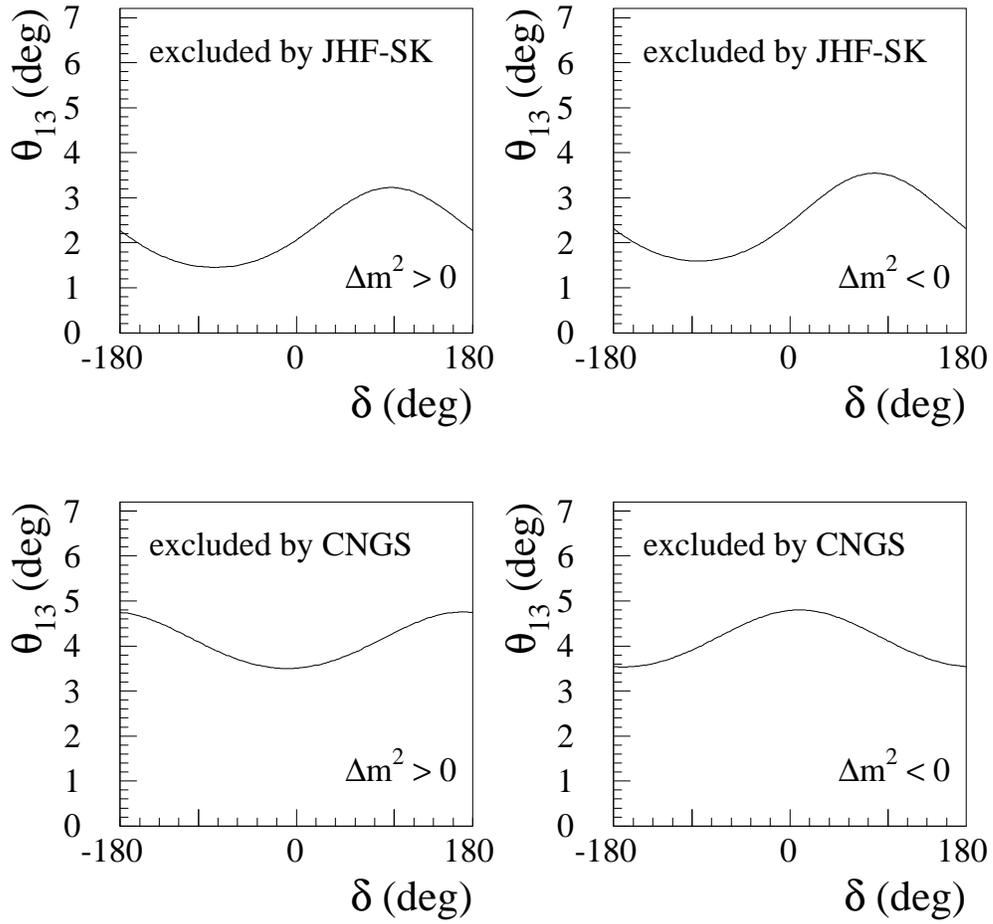,width=15cm}
\caption{$\sin^2 2\theta_{13}$ sensitivity at 90\% CL versus
$\delta$}
\label{fig:excl_2d}
\end{figure}

\begin{figure}[htbp]
\centering
\epsfig{file=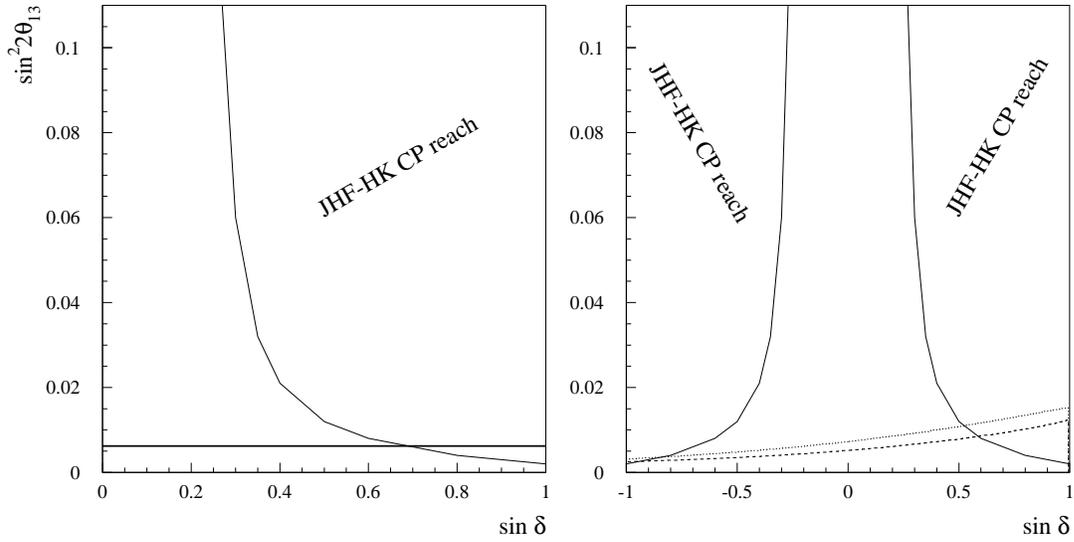,width=15cm}
\caption{Sensitivity to CP violation ($3\sigma$ discovery)
of JHF-HK as a function of $\sin \delta$ and 
$\sin^2 2\theta_{13}$. The horizontal band in (a) represent the value
excluded by JHF-SK at 90\% CL assuming 
$P(\nu_\mu \rightarrow \nu_e)\simeq O_1$. The corresponding exclusion
region for the full oscillation probability is shown in (b) for positive
(dashed line) or negative (dotted line) $\Delta m^2_{31}$.}
\label{fig:jhfhk}
\end{figure}

\noindent Exclusion plots for $\sin^2 2\theta_{13}$ are available
\cite{ICARUS,JHF,numioa,komatsu,diwan} in the approximation  
$P(\nu_\mu \rightarrow \nu_e)\simeq O_1$. This corresponds to the 
assumption $\delta=0$ for on-peak experiment ($\Delta \simeq \pi/2$)
and $\delta=\pi/2$ for off-peak ones ($\Delta \ll \pi/2$).
Hence, it is possible to extract the minimum accessible probability
$P_{min}$:

\begin{equation}
P_{min} \equiv \sqrt{\zeta}  \left\{ \frac{\beta}{Y\gamma^2} + 
\frac{\eta^2 \beta^2}{\gamma^2} \right\} ^{1/2}
\label{equ:pmin}
\end{equation}

\noindent from literature and compute Eq.~\ref{equ:final_inequality}
using the correct oscillation probability
\footnote{The results shown in this sections have been obtained
using the complete three family oscillation formula at constant matter
density and not the approximate expansion of Eq.~\ref{equ:probmatter}.}. 
Fig.~\ref{fig:excl_2d} shows the $\sin^2 2\theta_{13}$ sensitivity at
90\%~CL as a function of $\delta$ for JHF-SK and CNGS. 
Note that for positive values
of the CP phase, the $\delta$ dependence of JHF-SK 
has the worst possible behaviour
for a Phase~I experiment, since the minimum sensitivity 
to $\sin^2 2\theta_{13}$
is achieved at maximum CP violation (maximum discovery potential
of phase~II setups). This is illustrated in Fig.\ref{fig:jhfhk}.
Plot (a) shows the region where a $3\sigma$ discovery of CP violation 
at JHF-HK can be achieved as a function of $\sin \delta$ and 
$\sin^2 2\theta_{13}$ \cite{talk_nakaya}. The horizontal band is the exclusion
limit of JHF-SK at 90\%CL in the approximation 
$P(\nu_\mu \rightarrow \nu_e)\simeq O_1$. The correct exclusion limit
from JHF-SK is shown in plot (b) for positive (dashed line) or negative
(dotted line) $\Delta m^2_{31}$.
\noindent Assuming complete ignorance on the value of the CP phase and using
no other external information to lift the $\theta_{13}-\delta$
ambiguity (e.g. an $\bar{\nu}$ run with similar statistics), 
the actual excluded $\sin^2 2\theta_{13}$ is

\begin{equation}
\left. \sin^2 2\theta_{13} \right|_{excl} \ = \ \max_{-\pi < \delta < \pi} 
\left. \sin^2 2\theta_{13} \right|_{excl} (\delta)
\label{equ:real_excluded}
\end{equation}  

\noindent
In other words, the effective 
sensitivity is computed ``finding the largest value
of $\sin^2 2\theta_{13}$ which fits the true $\sin^2 2\theta_{13}=0$
at the selected confidence level'' \cite{lindner}.
If we assume complete ignorance also on the sign of $\Delta m^2_{31}$
the final excluded value of $\sin^2 2\theta_{13}$ is the maximum between the
value of $\sin^2 2\theta_{13}$ calculated by Eq.~\ref{equ:real_excluded}
assuming $\Delta m^2_{31}>0$ and the one with $\Delta m^2_{31}<0$.

\begin{figure}[htbp]
\centering
\epsfig{file=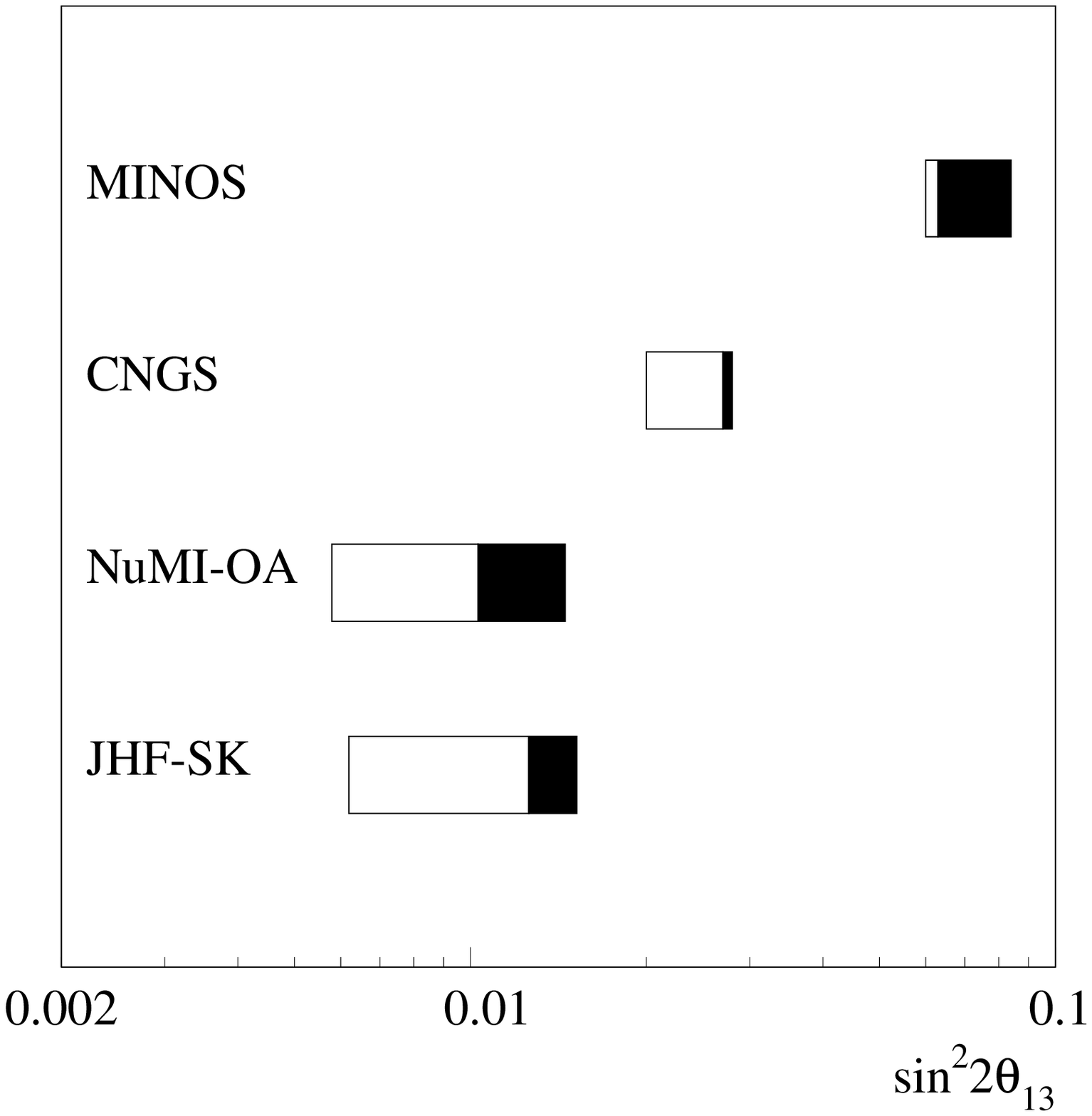,width=10cm}
\caption{$\sin^2 2\theta_{13}$ sensitivity at 90\% CL. Empty boxes
correspond to the deterioration due to the ignorance on the $\delta$
phase. Full boxes indicate further deterioration coming from the
sign of $\Delta m^2_{31}$.}
\label{fig:precloss}
\end{figure}

Fig.~\ref{fig:precloss} shows the expected precision for the
experiments considered in Sec.~\ref{osc_prob} for $|\Delta
m^2_{31}|=3\ 10^{-3}$~eV$^2$ and $\Delta m^2_{21}=7.3\
10^{-5}$~eV$^2$.  The empty boxes indicate the deterioration coming
from the integration on the CP phase $\delta$.  Full boxes show the
effect of the sign$(\Delta m^2_{31})$ degeneracy. A few comments are
in order.  Both JHF-SK and CNGS appear to be almost insensitive to the
sign of $\Delta m^2_{31}$ but in fact the effect is subtler.  In
JHF-SK, the leading term $O_1$ is independent of the transformation
$\Delta m^2_{31} \rightarrow - \Delta m^2_{31}$ thanks to the
smallness of the $\hat{A}$ parameter. $O_2$ is invariant under this
transformation and $O_3$ is suppressed as well as $O_4$. On the other
hand, at CNGS the leading term $O_1$ does not depend significantly on
the $\Delta m^2_{31}$ sign thanks to the cancellation of matter effect
at work for small values of $|(1-\hat{A})\Delta|$ (see
Sec.~\ref{osc_prob}).  The next-to-leading term in the oscillation
probability ($O_3$) is odd under the sign exchange. This effect is
equivalent to a $\delta \rightarrow \pi- \delta$ transformation so
that the same variation of probability appears during the integration
in $\delta$; hence, in Fig.~\ref{fig:precloss} the deterioration of
the sensitivity coming from the sign degeneracy is absorbed into the
deterioration caused by the CP phase.  This different behavior is
unveiled examining the exclusion plots at different values of $\delta$
(Fig.~\ref{fig:excl_2d})~\footnote{Note that for values of $\sin^2
2\theta_{13}$ close to the CHOOZ limit, it could be possible to use
synergically JHF-SK and CNGS to get information on the hierarchy of
neutrino masses.}. On the other hands, MINOS and NuMI~Off-Axis have
the highest sensitivity to the $\Delta m^2_{31} \rightarrow - \Delta
m^2_{31}$ transformation since the condition $\hat{A}\ne 0$ affects
directly the leading term $O_1$.  Plots similar to
Fig.~\ref{fig:precloss} have already been obtained for JHF and NuMI
Off-Axis by the authors of \cite{lindner} using a detailed simulation
of the setups. The bands of Fig.~\ref{fig:precloss} corresponding to
these experiments are in good agreement with their results. The CNGS
sensitivity has been cross-checked applying the full oscillation
probability to the analysis described in \cite{komatsu}. The
sensitivity deterioration at $|\Delta m^2_{31}|=3 \ 10^{-3}$~eV$^2$
for the two experiments separately is $0.025 \rightarrow
0.034$ (ICARUS) and $0.035 \rightarrow 0.045$ (OPERA).

\begin{figure}[htbp]
\centering
\epsfig{file=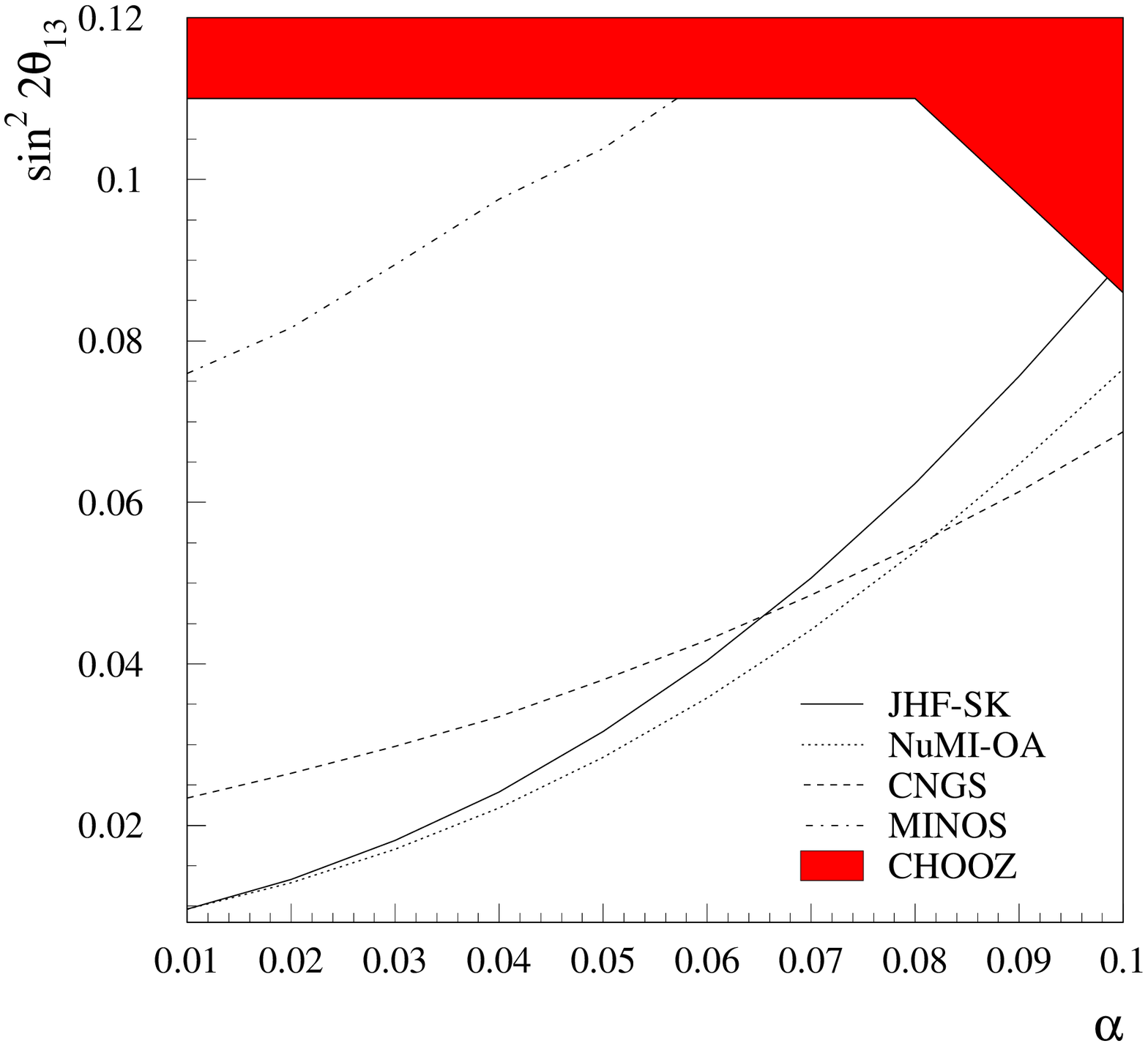,width=10cm}
\caption{$\sin^2 2\theta_{13}$ sensitivity at 90\% CL versus $\alpha 
\equiv \Delta m^2_{21}/ | \Delta m^2_{31} |$} 
\label{fig:alphadip}
\end{figure}

In Sec.~\ref{osc_prob} we argued that the trade-off between 
maximal $\sin^2 2\theta_{13}$ sensitivity and maximal PMNS reach
is connected with the size of the ratio 
$\alpha \equiv \Delta m^2_{21}/ | \Delta m^2_{31} |$. 
Fig.~\ref{fig:alphadip} shows the $\sin^2 2\theta_{13}$ sensitivity 
versus $\alpha$ for mass ratios up to $10^{-1}$. As expected,
the Phase~I experiments loose their capability to perform a ``pure''
$\sin^2 2\theta_{13}$ measurement in the high-LMA region of 
$\Delta m^2_{21}$. Note also that the present CHOOZ limits become
more stringent in the high-$\Delta m^2_{21}$ regime \cite{chooz_3family}.

\begin{figure}[htbp]
\centering
\epsfig{file=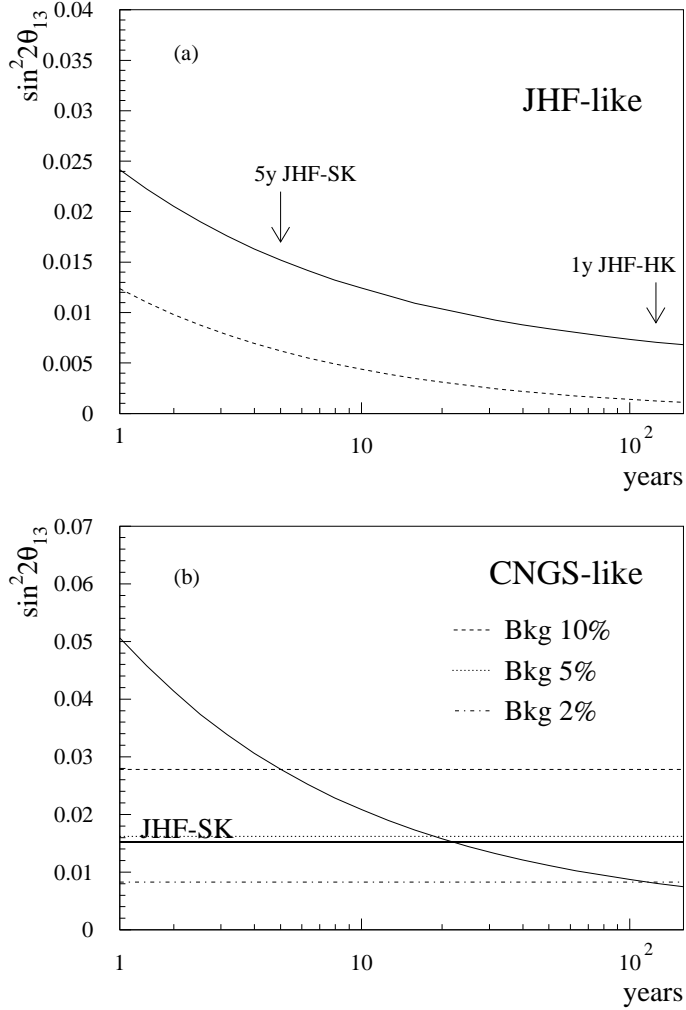,width=10cm}
\caption{$\sin^2 2\theta_{13}$ sensitivity at 90\% CL versus years
of exposures for JHF (a) and CNGS (b). In (a) the solid line represents
the sensitivity keeping into account the CP phase and 
sign$(\Delta m^2_{31})$ deterioration. The dashed line shows the
expected sensitivity assuming $P(\nu_\mu \rightarrow \nu_e)=O_1$.
In (b) the horizontal lines indicate the region where the 
$\sin^2 2\theta_{13}$ sensitivity becomes limited by the beam systematics
(see text).
} 
\label{fig:years}
\end{figure}

Fig.~\ref{fig:years}-a describes the sensitivity in $\sin^2
2\theta_{13}$ versus the integrated flux expressed in years of data
taking, assuming the nominal intensity of JHF-SK~\footnote{It
  corresponds to a proton intensity of 0.7~MW.  Note that 1 year of
  JHF-HK data taking corresponds to about 125 years of JHF-SK due to
  the increase of beam intensity and detector mass.}.  The limits have
been extracted rescaling naively with $\sqrt{Y}$ the minimum
accessible probability $P_{min}$ and ignoring the saturation effect
coming from the background normalization.  In fact, JHF is expected to
be limited by systematics only in the Phase~II of its physics
programme.  It is worth noting that the deterioration coming from the
degeneracies does not imply a plateau of the sensitivity. Phase~II
experiments will access a region of $\sin^2 2\theta_{13}$ deeper than
the one accessible by their Phase~I counterparts. As an example in
Fig.~\ref{fig:years}-a the achievable sensitivity on $\sin^2
2\theta_{13}$ after one year data taking of JHF-HK is shown ($\Delta
m^2_{21} = 7.3\times10^{-5}$~eV$^2$).

After the $\bar{\nu}$ run, JHF-HK will be able to observe maximal CP
violation in the leptonic sector down to $\sin^2 2\theta_{13} \sim 2
\times 10^{-3}$ for $\Delta m^2_{21} \sim 5 \times 10^{-5}$~eV$^2$
\cite{lindner_nufact} and the highest the solar mass, the better the
CP-sensitivity (the worse the Phase~I ``pure'' $\sin^2 2\theta_{13}$
sensitivity). So, a null result of JHF-SK cannot rule out convincingly
the possibility to perform PMNS precision physics with terrestrial
experiments.  Of course, this holds also for the Neutrino Factories which
have an even higher CP sensitivity than JHF-HK.  

Finally, Fig.~\ref{fig:years}-b shows the $\sin^2 2\theta_{13}$
sensitivity versus the exposure for a CNGS-like beam. For the actual
CNGS, the background systematics $\eta$ cannot be neglected. The
horizontal lines in the plot indicate the region where the beam
systematics will saturate the limits on $\sin^2 2\theta_{13}$
($\sqrt{B}=\eta B$).  They correspond to a precision in the
normalization of the $\nu_e$ background of 10\% and 5\%. The limit
from beam systematics for a setup with a near detector ($\eta \simeq
2$\%) is also shown.

\begin{figure}[htbp]
\centering
\epsfig{file=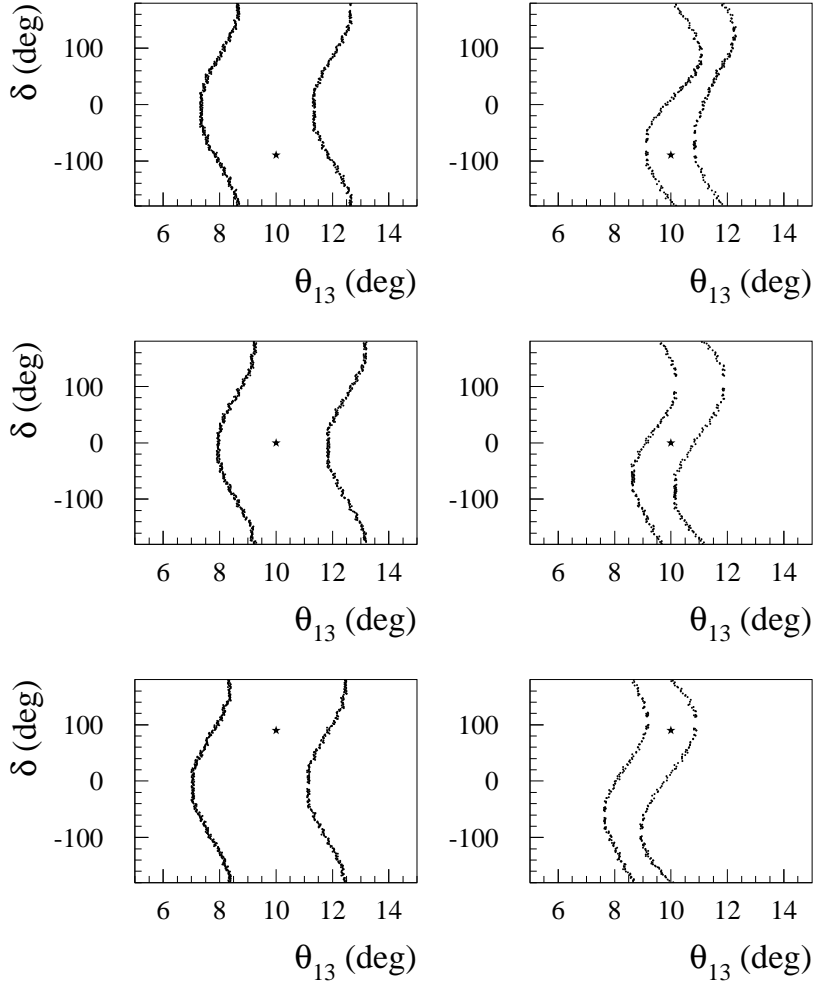,width=12cm}
\caption{Left plots: 90\%~CL allowed region after 5 years of CNGS data
taking  for $\theta_{13}=10^\circ$, normal hierarchy and
$\delta=-90^\circ$ (upper), $\delta=0^\circ$ (middle),
$\delta=90^\circ$ (lower plot). 
The plots on the right show the corresponding regions
obtained combining CNGS data with a 5-year run of JHF-SK.
The stars indicate the true value of $\theta_{13}$ and $\delta$.}
\label{fig:discovery}
\end{figure}

Up to now, we only considered the interplay between on-peak and
off-peak beams in the case of null result. However, for high values of
$\theta_{13}$ ($\theta_{13} \simge 7^\circ$) CNGS could be able to
establish $\nu_\mu \rightarrow \nu_e$ oscillations at $3\sigma$ level
for any value of $\delta$. In this scenario, a very strong improvement
in the measurement of the angle (as a function of $\delta$) is
obtained after the JHF-SK data taking. The three plots on the left of
Fig.~\ref{fig:discovery} show the 90\%~CL allowed region after 5 years
of CNGS data taking for $\theta_{13}=10^\circ$, normal hierarchy and
$\delta=-90^\circ$ (upper), $\delta=0^\circ$ (middle),
$\delta=90^\circ$ (lower plot). The plots on the right show the
corresponding regions obtained combining CNGS data with a 5-year run
of JHF-SK.  Note that the combined $(\theta_{13},\delta)$ band has no
more uniform width, as it would be for JHF-SK alone, and shrinking of
the region around $\delta=\pm 90^\circ$ results from the combination
of experiments with different $(\theta_{13},\delta)$ patterns.
Clearly, it is possible to lift explicitly the $(\theta_{13},\delta)$
correlation  after a $\bar{\nu}$ run. For the optimization of the
JHF-SK $\nu+\bar{\nu}$ data taking in case of positive signal, we
refer to~\cite{kajita}.

\section{Conclusions}
\label{conclusions}

Phase~I experiments will measure the parameters leading the oscillations
at the atmospheric scale with unprecedented precision. They will fix the
$\sin^2 2\theta_{23}$ and $|\Delta m^2_{31}|$ terms at the 1\% level and
observe a clear oscillation patters in $\nu_\mu$ disappearance mode.
Moreover, they can test the subdominant $\nu_\mu \rightarrow \nu_e$ transition 
improving significantly the present knowledge of $\theta_{13}$. On the other
hand, the actual sensitivity to $\sin^2 2\theta_{13}$ is  strongly deteriorated
by the present ignorance on the CP violating phase and the sign of 
$\Delta m^2_{31}$. In Sec.~\ref{numerical_calc} it has been shown
that, in case of null result,  the improvements in the
exclusion limits for $\sin^2 2\theta_{13}$  will be marginal with respect
to long baseline experiments like ICARUS and OPERA
($0.03 \rightarrow 0.015$) at $\alpha \simeq 0.02$ and
negligible for higher values of $\alpha$. On the other hand, a high
solar scale ($\alpha>0.02$) enhances significantly the 
capability of Superbeam and Neutrino Factory
to access CP violation even for values of $\sin^2 2\theta_{13} 
\sim {\cal O}(10^{-3}\div 10^{-4})$. 
Hence, a null result at Phase~I will not constrain in a significant way the
physics reach of SB/NF. Clearly, it is impossible to tune a Phase~I
experiment to reach simultaneously a high $\sin^2 2\theta_{13}$ sensitivity
(setups ``ancillary'' to Phase~II) and a high sensitivity to the CP phase
and the mass hierarchy (setups with high ``PMNS reach''). At present
we do not know if JHF-SK and NuMI~Off-Axis belong to the former or latter
category, due to the large uncertainty on $\alpha$. Anyway, a real 
Phase~I experiment (or cluster of experiments) performing a ``pure''
$\sin^2 2\theta_{13}$ along the line proposed by the authors of
\cite{lindner,kajita,minakata} would be highly advisable to firmly ground
the SB/NF physics programme.  

\section*{Acknowledgements}
We thank A.~Donini, M.~Lindner, M.~Mezzetto and P.~Strolin for discussions and 
careful reading of the manuscript. 
We are indebted to D.~Meloni for providing us with the 
three family oscillation code and to P.~Huber for useful remarks
on Eq.~\ref{equ:probmatter}.


\end{document}